\pdfoutput=1

\documentclass[aps,prl,10pt,twocolumn,showpacs,superscriptaddress]{revtex4-1}

\usepackage{latexsym}
\usepackage{graphics,epstopdf}
\usepackage{newlfont}
\usepackage{amssymb}
\usepackage{amsfonts}
\usepackage{amsmath}
\usepackage{amsthm}
\usepackage{graphicx}
\usepackage{epsfig}
\usepackage{color}
\usepackage{bm}
\usepackage{times}

\begin{document}
\title{Weak measurement induced super discord can resurrect lost quantumness}
\author{Uttam Singh}
\author{Arun Kumar Pati}
\affiliation{Quantum Information and Computation Group\\
Harish-Chandra Research Institute, Chhatnag Road, Jhunsi, 
Allahabad 211 019, India}

\date{\today}

\begin{abstract}
The projective measurement usually destroys the quantum correlation between two subsystems of a 
composite system, thereby
making the measured state useless for any efficient quantum information processing and quantum computation task. The weak
measurement acts gently on quantum system and do not force the system state to decohere completely, thus revealing the super
quantum discord, which can take value more than that of the normal quantum discord.
%We show that the measured state is not completely useless as it has non zero super quantum discord, which is a recently proposed new
%measure of quantum correlation.
% and is weak measurement generalization of the quantum discord. 
Remarkably, we prove that the super quantum discord in the post measured state is equal to the difference between the super
quantum discord and the normal quantum discord in the original state. Thus, the weak measurement has the ability to resurrect
the lost quantumness of any composite quantum state.
%i.e., $\tilde{D}_w(A:B) = D_w(A:B) - D_s(A:B)$.
%The extra quantumness, which is captured by super quantum discord and is
%quantified as the difference between the SQD and the normal quantum discord, is a function of state only and is independent of
%measurements. So it may be stated as a conservation law for extra quantum correlation in the state. This conservation law, if
%exists, can help in clarifying the issues of creation and destruction of correlations in a quantum system.
% The findings of this paper can be used to harness the quantum resources efficiently.
This suggests a conservation law for the extra quantum correlation in any composite state. The amount of extra quantum
correlation which is destroyed by the projective measurement in the original state is equal to the amount of extra
quantum correlation captured by the weak measurement in the post-measured state.

\end{abstract}

\maketitle

\emph{Introduction.}--
Quantum measurement is one of the central concepts in the theory of quantum information and computation. In quantum theory,
quantum measurement plays a distinct role.
%are peculiar in many regards.  Among their many peculiarities one which often come across in almost every quantum information 
%processing and computation task is the fact that 
Unlike the unitary dynamics, the measurement process inevitably disturbs the quantum state 
unless the state is in one of the eigenstates of the measured
observable \cite{von}. This is in contrast with the measurement process in the classical world 
where it does not affect the state of the
classical system. The quantum measurement process destroys the information stored in the original quantum state.
As a result, we cannot measure a single quantum system to know its state. This property of quantum
measurement process comes handy in many security protocols where it can help in detecting 
the eavesdropping \cite{ben}.
Moreover, this can be used to freeze the dynamics of quantum systems which is sometimes desired and quantum zeno effect is
the prime example of it \cite{baidya, itano}.
%But at the same time this also creates hurdle for us as we can not measure a single quantum system to know its state.
If we consider local projective measurements on a bipartite system, in effect they remove the quantum correlation between two parties
and make them classical. But most of the time we
require to maintain the quantum correlations as they are believed to be very precious resources for almost all better than classical
information processing and computation tasks. Therefore, it is desirable to have a handle on the measurement process
to maintain correlations during the measurement process or to have a scheme for 
the resurrection of quantum correlation.

Typically, we encounter projective and positive operator valued measurements in quantum mechanics. However
Aharonov, Albert, and Vaidman have introduced the notion of weak measurements \cite{aav}, which
actually has control on the strength of the measurement process and it was introduced for pre and post selected ensembles.
Later on, it was generalized by Oreshkov and Brun \cite{brun} to the case of preselection only and there it was proved that
weak measurements are universal as any projective or positive operator valued measurement can be realized as a set of sequential
weak measurements. These are the most general ones as any measurement can be written as a special case of weak measurements. The
weak measurement has found several applications like resolving Hardy's non-locality \cite{js}, amplification of the deflection
of optical beam in the Sagnac interferometer \cite{dix}, direct measurement of wave function of single photon \cite{jsl},
understanding and developing the theory of entanglement monotones \cite{ores,chit}, protection of quantum states and at the
same time allowing for universal quantum computation or quantum control \cite{lidar}, and protecting quantum entanglement from
decoherence \cite{kim}.

Quantum correlation is the strongly believed reason for any efficient quantum information processing and computing task compared
to their classical counterparts. So it is necessary to quantify the correlations in quantum systems in order to harness them
efficiently for useful tasks. Entanglement, which is one of the measures of quantum correlation, has played important role in
many quantum information protocols. Quantum discord is yet another celebrated measure of quantum correlation and
it was introduced by Ollivier and Zurek \cite{zurek}. The notion of quantum discord has attracted a lot of attention in the 
recent years \cite{mera,modi}.
The quantum discord represents the locally inaccessible information. It is
the difference between the total and the classical correlation \cite{hend}. The quantum discord can be nonzero even for
some separable states, where entanglement is zero. This tells us that quantum discord may capture the quantum correlation for mixed
states that is left undetected by the entanglement. In the absence of quantum entanglement, the quantum discord is one of most
believed reasons in any speed up in quantum computation or in any efficient information processing task. This is exemplified
from the possibility of giving the power to quantum computation in the absence of entanglement \cite{animesh}, quantum
communication such as quantum state merging \cite{madhok,caval}, quantum entanglement distribution with separable states
\cite{alex,chuan} and has also been experimentally used as a resource for remote state preparation 
\cite{akp,dak,gu}. Quantum discord has been investigated
in finding conditions for the monogamy nature of quantum correlations \cite{prabhu,gio}. It has been shown that the quantum
discord is also a physical quantity, because erasure of quantum correlation must lead to entropy production in the system
and the environment \cite{akp}.

The weak measurements offer a flexible way of interrogating a composite quantum system and at the same time maintaining quantum
correlation between the parts of the system. Recently, it has been shown that weak measurement can reveal more quantum
correlation \cite{utm} in a composite state as it captures the otherwise destroyed quantum correlation. This fact has been
shown using the notion of super quantum discord (SQD) as a measure of quantum correlation, which is weak measurement
generalization of the quantum discord. The remarkable property of the SQD is that it is more than one for the maximally entangled
state, which can never be anticipated using projective measurements. The SQD is a completely new tool in our arsenal to
detect and understand quantum correlation in composite systems. It shows that the quantum correlation in a composite state
is not only an observer dependent notion but also depends on how strongly or gently one performs measurement on the system.
It has also been shown that the SQD is always greater than or equal to the normal discord and nonzero even if the normal
discord is zero. The maximum value that SQD can reach is equal to the quantum mutual information in the system. This gives
a support to the evidences that total correlation behaves as if it is exclusively quantum 
\cite{bennett}. In  a recent paper, the super discord
has been found to play a role in the protocol of optimal assisted state discrimination 
\cite{boli}.  It has been used to investigate the monogamy nature of quantum states \cite{mhu}.

Imagine that we have a bipartite system in a state $\rho_{AB}$ on which we perform a 
projective measurement on the subsystem $B$. The amount of inaccessible information or  
the quantum correlation is captured by the normal discord $D(\rho_{AB})$. The state of the 
composite system after the projective 
measurement is given by $\tilde\rho_{AB} = \sum_ip_i\rho_{A|i}\otimes |i\rangle\langle i|$. 
However, instead of projective measurement had we performed the weak measurement, we could have 
revealed the super quantum discord. Therefore, in a sense, by performing the projective measurement
we have destroyed the extra quantum correlation.
In this paper we will address the 
question, can there be a procedure which can resurrect the extra quantum correlation that is lost due to a complete local projective
measurement on one of the subsystems of composite system? Here, we will show that if we consider the SQD as a measure of quantum
correlation in the composite system, it is nonzero in the state which we get after the complete local projective measurement.
Moreover, the SQD in the post-measured state is equal to the difference between the SQD and the normal quantum discord in the original
state. Thus, the weak measurement induced super quantum discord can resurrect the lost quantum correlation of the original state
for any given measurement strength. This demonstrates that the extra quantum correlation is a conserved quantity and 
it depends only on the state of the system and the measurement strength. In essence, even if an initial projective measurement cannot\
capture the extra quantumness, later on by performing the weak measurement we can reassure that 
indeed there was this extra 
quantum correlation (had we performed a weak measurement before). This can also be interpreted as 
delayed revelation of extra quantum correlation using the weak measurement.\\

{\it Weak measurements and the super quantum discord}.--
The concept of weak measurements can be formulated using the measurement operator formalism \cite{brun}. The weak measurement
operators are given by 
%$P(x) = \sqrt{\frac{(1-\tanh x)}{2}} \Pi_0 + \sqrt{ \frac{(1+\tanh x)}{2}} \Pi_1$ ,
% $P(-x) = \sqrt{\frac{(1+\tanh x)}{2}}\Pi_0 + \sqrt{\frac{(1-\tanh x)}{2}}\Pi_1$ ,
\begin{eqnarray}
 P(x) &=& \sqrt{\frac{(1-\tanh x)}{2}} \Pi_0 + \sqrt{ \frac{(1+\tanh x)}{2}} \Pi_1,  \nonumber\\
 P(-x) & = & \sqrt{\frac{(1+\tanh x)}{2}}\Pi_0 + \sqrt{\frac{(1-\tanh x)}{2}}\Pi_1, \nonumber\\
\end{eqnarray}
where $x$ is a parameter that denotes the strength of the measurement process, $\Pi_0$ and $\Pi_1$ are two orthogonal projectors
with $\Pi_0 + \Pi_1 =I$. The weak measurement operators satisfy $P^{\dagger}(x)P(x) + P^{\dagger}(-x)P(-x) = I$. These operators
have the following properties: (i) $P(0) = \frac{I}{\sqrt 2}$ resulting in no state change, (ii) in the strong measurement limit
we have the projective measurement operators, i.e., $\lim_{x \rightarrow \infty} P(-x) = \Pi_0$ and
$\lim_{x \rightarrow \infty} P(x) = \Pi_1$, (iii) $ P(x) P(y)\propto P(x+y)$, and (iv) $[P(x), P(-x)] =0$. Now consider a bipartite
state $\rho_{AB}$. After we perform weak measurement on the subsystem $B$ by the weak operators $\{P^{B}(x),P^{B}(- x)\}$, the
post-measurement state for the subsystem $A$ is given by 
\begin{equation}
\rho_{A|P^{B}(\pm x)}=\frac{\mbox{Tr}_{B}[(I \otimes P^{B}(\pm x)) \rho_{AB} (I \otimes  P^{B}(\pm x))]} 
{\mbox{Tr}_{AB}[(I \otimes P^{B}(\pm x)) \rho_{AB} (I \otimes P^{B}(\pm x))]}
\end{equation}
and the probability with which this occurs is given by 
\begin{equation}
 p(\pm x) = 
\mbox{Tr}_{AB}[(I \otimes P^{B}(\pm x)) \rho_{AB} (I \otimes P^{B}(\pm x))].
\end{equation}
The ``weak quantum conditional entropy'' after we perform a weak measurement on the subsystem $B$, is given by
\begin{equation}
S_w(A|\{P^{B}(x)\})= p(x) S(\rho_{A|P^{B}(x)}) + p(-x) S(\rho_{A|P^{B}(-x)}).
\end{equation}
The SQD in the state $\rho_{AB}$, denoted by $D_w(\rho_{AB})$, is defined as \cite{utm}
\begin{equation}
D_w(\rho_{AB}):=  \min_{\{\Pi_i^B\}}  S_w(A|\{P^{B}(x)\}) - S(A|B),
\end{equation}
where $ S(A|B)= S(\rho_{AB}) - S(\rho_{B})$. The SQD satisfies $I(\rho_{AB})\geq D_w(\rho_{AB})\geq D_s(\rho_{AB})$. It is a monotonic
function of the measurement strength $x$ \cite{utm}. The extra quantum correlation captured by the weak measurement is defined as
$\Delta(\rho_{AB}) = D_w(\rho_{AB}) - D_s(\rho_{AB})$. In the strong measurement limit, the extra quantum correlation becomes zero,
i.e., $\lim_{x\rightarrow\infty}\Delta(\rho_{AB}) = 0$.
The novel feature of the weak measurement is that it is able to capture the 
extra quantum correlation in the original state.% as well as in the post-measured state.
%Now we will present the main result of this paper, which shows that SQD is non zero
%in the measured state, which is obtained after the local projective measurement performed on the subsystem $B$.

%\noindent {\bf Theorem}: 
Below we will prove that given a bipartite state $\rho_{AB}$, the difference between the super quantum discord (SQD) revealed by the weak measurement
and the normal discord in the initial state is equal to super discord in the state after strong measurement on the initial state,
i.e., $D_w(\rho_{AB}) - D_s(\rho_{AB}) = D_w(\tilde\rho_{AB})$.\\
%if and only if weak conditional entropy of initial state and weak conditional entropy for measured new state are minimized in
%same basis.
%\noindent {\bf Proof:} 
Consider a bipartite state $\rho_{AB}$ and suppose that we perform a projective measurement on 
the subsystem $B$.
Without loss of generality let us assume that conditional entropy of $\rho_{AB}$ is
minimized for the orthogonal projectors $\{\Pi_\psi^{B},\Pi^B_{\bar{\psi}}\}$, where $\Pi_\psi^{B}=|\psi\rangle\langle\psi|$ and 
$\Pi_{\bar\psi}^B=|\bar\psi\rangle\langle\bar\psi|$. The state of the subsystem $A$, after the application of the measurement operator $\Pi_\psi^{B}$ becomes
\begin{align}
\rho^A_1=\frac{\mbox{Tr}_{B}[(I \otimes \Pi_\psi^{B}) \rho_{AB} (I \otimes  \Pi_\psi^{B})]}{p_1}
=\frac{_B\langle\psi|\rho_{AB}|\psi\rangle_B}{p_1}
\end{align}
and it occurs with probability $p_1=\mbox{Tr}_{AB}[(I \otimes \Pi_\psi^{B}) \rho_{AB} (I \otimes \Pi_\psi^{B})]$.
Similarly, the state of the subsystem $A$, after the application of the measurement operator $\Pi^B_{\bar{\psi}}$ becomes
\begin{align}
\rho^A_2=\frac{\mbox{Tr}_{B}[(I \otimes \Pi^B_{\bar{\psi}}) \rho_{AB} (I \otimes  \Pi^B_{\bar{\psi}})]}{p_2}
=\frac{_B\langle{\bar{\psi}}|\rho_{AB}|{\bar{\psi}}\rangle_B}{p_2}
\end{align}
and it occurs with probability $p_2=\mbox{Tr}_{AB}[(I \otimes \Pi^B_{\bar{\psi}}) \rho_{AB} (I \otimes \Pi^B_{\bar{\psi}})]$.
The normal discord is given by
\begin{align}
\label{strong}
 D_s(\rho_{AB}) =S(A|\{\Pi_\psi^{B}\}) - S(A|B),
\end{align}
where $S(A|\{\Pi_\psi^{B}\})=p_1S(\rho^A_1)+p_2S(\rho^A_2)$. 
Now let us calculate the super quantum discord for state $\rho_{AB}$.
Let the weak measurement
operators $\{P^B(x), P^B(-x)\}$, that are given by $P^B(\pm x)= 
a(\pm x)\Pi^B_{\psi}+a(\mp x)\Pi^B_{\bar{\psi}}$, where $a(\pm x)=
\sqrt{\frac{1\mp\tanh x}{2}}$ and $\{|\psi\rangle,|\bar{\psi}\rangle\}$ forms an orthonormal 
basis, will minimize the weak conditional entropy.
Given a general density matrix $\rho_{AB}$,
%and let us assume that the weak operators $\{P^B(x), P^B(-x)\}$, with  
%$P^B(\pm x)=a(\pm x)\Pi^B_{\psi}+a(\mp x)\Pi^B_{\bar{\psi}}$ minimize the weak conditional entropy.
it has been proved that the weak conditional entropy is a monotonic and continuous function of 
the measurement strength \cite{utm}. 
The continuity of the weak conditional entropy with $x$, then implies that in the limit 
$x\rightarrow \infty$, $\{\Pi^B_{\psi}, \Pi^B_{\bar{\psi}}\}$ should minimize the strong 
conditional entropy. To understand this differently, 
let us assume that the weak conditional entropy and the strong conditional entropy are 
minimized with two different bases. If it is so,
then under the strong measurement limit, we will obtain two distinct values for the normal 
quantum discord, and this cannot happen. 
Therefore, the measurement basis which minimizes the strong conditional entropy is same as the 
basis that minimizes the weak conditional entropy.
%For a general bipartite density matrix, where the conditional entropy depends on all parameters of the basis (for qubit case,
%two parameters), the weak conditional entropy becomes the normal conditional entropy in $x\rightarrow \infty$ limit. Therefore, the basis which
%minimizes the normal conditional entropy should also minimize the weak conditional entropy. 

%Also the  weak conditional entropy is minimized in this basis.
Now the conditional state of subsystem $A$ after the measurement of weak operators on subsystem $B$ is given by
\begin{align}
 \rho_{A|P^{B}(\pm x)} &=\frac{\mbox{Tr}_{B}[(I \otimes P^{B}(\pm x)) \rho_{AB} (I \otimes  P^{B}(\pm x))]}{p(\pm x)}\nonumber \\
 &=\frac{a(\pm x)^2p_1\rho^A_1+a(\mp x)^2p_2\rho^A_2}{p(\pm x)},
\end{align}
where $p(\pm x)=\mbox{Tr}_{AB}[(I \otimes P^{B}(\pm x)) \rho_{AB} (I \otimes P^{B}(\pm x))]=a(\pm x)^2p_1+a(\mp x)^2p_2$.
The super quantum discord is given by
\begin{align}
\label{weak}
 D_w(\rho_{AB}) = S_w(A|\{P^{B}(x)\}) - S(A|B),
\end{align}
where $S_w(A|\{P^{B}(x)\})=p(x)S(\rho_{A|P^{B}(x)})+p(-x)S(\rho_{A|P^{B}(-x)})$. The difference between the SQD and
the normal quantum discord is the extra quantumness, denoted by $\Delta$, and is given by
\begin{align}
\label{diff}
 \Delta :&= D_w(\rho_{AB}) - D_s(\rho_{AB})\nonumber \\ &=S_w(A|\{P^{B}(x)\})-S(A|\{\Pi_\psi^{B}\}).
\end{align}
This extra quantum correlation is a function of the state and the measurement strength. It is possible to reveal the extra quantum correlation
using the weak measurements only. But suppose that instead of weak measurement, we perform projective measurement on the subsystem $B$.
Then, obviously, we will loose the extra quantumness ( as $ \lim_{x\rightarrow\infty}\Delta(x)=0 $ ).
The state $\rho_{AB}$ after a measurement of $\{\Pi_\psi^{B},\Pi^B_{\bar{\psi}}\}$, becomes
\begin{align}
\label{new}
 \tilde{\rho}_{AB}=&(I \otimes \Pi_{\psi}) \rho_{AB} (I \otimes  \Pi_{\psi}) 
 + (I \otimes \Pi^B_{\bar{\psi}}) \rho_{AB} (I \otimes  \Pi^B_{\bar{\psi}})\nonumber \\
 &=p_1\rho^A_1\otimes|\psi\rangle\langle\psi|+p_2\rho^A_2\otimes|\bar{\psi}\rangle\langle\bar{\psi}|.
\end{align}
Now, quite surprisingly, we will show that if we perform weak measurement on the post-measured state $\tilde{\rho}_{AB}$, we can recover the extra quantumness.
Let us consider the weak measurement operators $\{Q^B_\phi(x),Q^B_\phi(-x)\}$ , where $Q^B_\phi(\pm x) = a(\pm x)\Pi_{\phi}
+a(\mp x)\Pi_{\bar{\phi}}$, where $a(\pm x)=\sqrt{\frac{1\mp\tanh x}{2}}$ and $\{|\phi\rangle,|\bar{\phi}\rangle\}$ forms
orthonormal basis. Also let us consider that the weak conditional entropy of the state $\tilde{\rho}_{AB}$ is minimized in this
basis. Let us denote the conditional states after the measurement of $Q^B_\phi(x)$, $Q^B_\phi(-x)$ by
$\tilde{\rho}_{A|Q^B_\phi(x)}$ and $\tilde{\rho}_{A|Q^B_\phi(-x)}$ and the corresponding probabilities are 
$q(x)$ and $q(-x)$, respectively. Now we have
\begin{align}
 &q(\pm x)\tilde{\rho}_{A|Q^B_\phi(\pm x)} = \mathrm{Tr}_{B}[(I \otimes Q^{B}(\pm x)) \tilde{\rho}_{AB}
 (I \otimes  Q^{B}(\pm x))]\nonumber \\
 &=a(\pm x)^2[p_1\rho^A_1|\langle\phi|\psi\rangle|^2+p_2\rho^A_2|\langle\phi|\bar{\psi}\rangle|^2]\nonumber \\
 &~~~~~~~~~~~~~~~+a(\mp x)^2[p_1\rho^A_1|\langle\bar{\phi}|\psi\rangle|^2+p_2\rho^A_2|\langle\bar{\phi}|\bar{\psi}\rangle|^2],
\end{align}
where $q(\pm x) = a(\pm x)^2[p_1|\langle\phi|\psi\rangle|^2+p_2|\langle\phi|\bar{\psi}\rangle|^2]
+a(\mp x)^2[p_1|\langle\bar{\phi}|\psi\rangle|^2+p_2|\langle\bar{\phi}|\bar{\psi}\rangle|^2]$. Therefore, the super discord in the state
$\tilde{\rho}_{AB}$ is given by
\begin{align}
\label{newweak}
 D_w(\tilde\rho_{AB}) %&= S(\tilde{\rho}_B) - S(\tilde{\rho}_{AB}) + q(x)S(\tilde{\rho}_{A|Q^B_\phi(x)})\nonumber \\
 %&~~~~~~~~~~~~~~~~~~~~~~~~~~~~~+ q(-x)S(\tilde{\rho}_{A|Q^B_\phi(- x)})\nonumber \\
 &=q(x)S(\tilde{\rho}_{A|Q^B_\phi(x)}) + q(-x)S(\tilde{\rho}_{A|Q^B_\phi(- x)})\nonumber \\
 &~~~~~~~~~~~~~~~~~~~~~~~~-[p_1S(\rho^A_1)+p_2S(\rho^A_2)].
\end{align}
Comparing (\ref{diff}) and (\ref{newweak}) we get that $D_w(\tilde\rho_{AB})$ and $\Delta$ are equal if and only if 
(i) $|\langle\phi|\psi\rangle|^2$ = 1 = $|\langle\bar{\phi}|\bar{\psi}\rangle|^2$ and
(ii) $|\langle\phi|\bar{\psi}\rangle|^2$ = 0 = $|\langle\bar{\phi}|\psi\rangle|^2$. In quantum mechanics the states
which differ by an overall phase are equivalent to each other. Below, we argue that indeed it is the case.

%Given a general density matrix $\rho_{AB}$, let us assume that the weak operators $\{P^B(x), P^B(-x)\}$, where 
%$P^B(\pm x)=a(\pm x)\Pi^B_{\psi}+a(\mp x)\Pi^B_{\bar{\psi}}$ minimize the weak conditional entropy.
%It has been proved that the weak conditional entropy is a monotonic and continuous function of the measurement strength \cite{utm}. 
%The continuity of the weak conditional entropy with $x$, then implies that in the limit $x\rightarrow \infty$, $\{\Pi^B_{\psi}, \Pi^B_{\bar{\psi}}\}$ should minimize the strong 
%conditional entropy. (Assume that the weak conditional entropy and the strong conditional entropy are minimized with two different bases. If it is so, then under the strong measurement limit, 
%we will obtain two distinct values for the strong conditional entropy, and this is a contradiction). Therefore, the measurement basis which minimizes the strong conditional entropy is
%same as the basis that minimizes the weak conditional entropy.

First, we note that the classical correlation for the state  $\tilde{\rho}_{AB}$ under the 
weak measurement
is given by $S(A) - S_w(A|\{Q^{B}(x)\})$. In the limit $x\rightarrow \infty$, this becomes 
the classical correlation for the strong measurement, 
i.e.,  $S(A) - S(A|\{\Pi^{B}_{\psi}\})$.
%\emph{$\Delta$ and $D_w(\tilde\rho_{AB})$ for mixed states.--}
Again, from the continuity of the conditional entropy, we must have $\{\Pi^B_{\psi}, \Pi^B_{\bar{\psi}}\} = 
\{\Pi^B_{\phi}, \Pi^B_{\bar{\phi}}\}$. 
%If the strong conditional entropy does not depend on all the
%parameters of measurement basis then there is redundancy and the two measurement bases may be different, in general. 
%(One needs to minimize the basis in the case of weak measurement separately.) 
This completes the proof of the main result of this paper. 

We emphasize that this is a completely counter intuitive quantum effect. The extra quantum correlation, which is present in the original state,
can be resurrected even after it is lost. The weak measurement has the ability to do so. The extra quantum correlation is shown
to be precisely equal to the super quantum discord in the post-measured state.
Now we provide some illustrative examples to felicitate our main claim.

\emph{The extra quantum correlation for general pure states.--}
For pure entangled state $|\psi\rangle_{AB}=\sqrt{\lambda_0}|00\rangle+\sqrt{\lambda_1}|11\rangle$, the strong conditional entropy is
given by $ S(A|\{\Pi^B_\psi\})=0$ and it is true in any basis. The normal discord for any pure bipartite state is equal to the entanglement
entropy, so $D_s(\rho_{AB}) = S(\rho_B) = -\lambda_0\log\lambda_0 -\lambda_1\log\lambda_1 $. The difference between the super quantum discord
and the normal discord in the original pure entangled state is given by \cite{utm} 
\begin{align}
\Delta = D_w(\rho_{AB}) &- D_s(\rho_{AB})=\min_\theta\big\{-\sum_{y=\pm x}p(y)[\nonumber \\
&k_+(y)\log k_+(y) + k_-(y)\log k_-(y)]\big\},
%-p(-x)[\nonumber\\
%&k_+(-x) \log k_+(-x) +k_-(-x)\log k_-(-x)],
\end{align}
where
$p(\pm x)= \frac{1}{2}{[1\mp(\lambda_0-\lambda_1)\tanh x \cos \theta]}$, $k_{\pm}(x)  =\frac{1}{2}[1\pm
\sqrt{1 - \frac{\lambda_0\lambda_1}{(p(x)^2\cosh^2{x})}}]$ and $k_{\pm }(-x) $ can be defined similarly. For $x=0.2$ and
$\lambda_0=0.2$, the minimum of weak conditional entropy occurs at $\theta=\pi/2$ and this corresponds to
$\{|+\rangle,|-\rangle\}$  basis. Therefore, the extra quantum correlation is given by $\Delta= 0.7010$.
Now the pure state after a measurement in $\{|+\rangle,|-\rangle\}$ basis is given by
\begin{align}
\tilde{\rho}_{AB} &=
%& \frac{1}{2}[\lambda_0|0\rangle\langle0|+\lambda_1|1\rangle\langle1|+\sqrt{ \lambda_0 \lambda_1}(|0\rangle\langle1|
%+|1\rangle\langle0|)]|+\rangle\langle+|\nonumber\\
%& \frac{1}{2}[\lambda_0|0\rangle\langle0|+\lambda_1|1\rangle\langle1|-\sqrt{ \lambda_0 \lambda_1}(|0\rangle\langle1|
%+|1\rangle\langle0|)]|-\rangle\langle-|\nonumber\\
 \frac{1}{2}[\lambda_0|0\rangle\langle0|+\lambda_1|1\rangle\langle1|]\otimes I_B\nonumber\\
&+\frac{\sqrt{ \lambda_0 \lambda_1}}{2}(|0\rangle\langle1|+|1\rangle\langle0|)]\otimes(|0\rangle\langle1|+|1\rangle\langle0|).
\end{align}
Here $S(\tilde{\rho}_{AB})= 1 =S(\tilde{\rho}_{B})$. The weak conditional entropy can be calculated using
$\{Q^B_\eta(x),Q^B_\eta(-x)\}$ , where $Q^B_\eta(\pm x) = a(\pm x)\Pi_{\eta}+a(\mp x)\Pi_{\bar{\eta}}$ with
$\{|\eta\rangle,|\bar{\eta}\rangle\}$ as an orthogonal basis. Here $|\eta\rangle=\cos\frac{\gamma}{2}|0\rangle
+\exp(i\delta)\sin\frac{\gamma}{2}|1\rangle$ and $|\bar\eta\rangle=\cos\frac{\gamma}{2}|1\rangle
-\exp(-i\delta)\sin\frac{\gamma}{2}|0\rangle$. The state of the system after measurement of $Q^B_\eta(x)$ is given by
\begin{align}
&\tilde{\rho}_{A|Q^B_\eta(x)}= \lambda_0|0\rangle\langle0|+\lambda_1|1\rangle\langle1| \nonumber\\
&- \sqrt{ \lambda_0 \lambda_1}\sin\gamma\cos\delta\tanh x(|0\rangle\langle1|+|1\rangle\langle0|),
\end{align}
and the probability of this outcome is $1/2$. Similarly, for $Q^B_\eta(-x)$ we have $q(-x)=1/2$. 
Thus, the weak conditional
entropy in the measured state is given by
\begin{align}
&S_w(A|\{Q_\eta^B(x)\}) =\min_{\{\gamma,\delta\}}[- (\frac{1+l(x,\gamma,\delta)}{2})\log(\frac{1+l(x,\gamma,\delta)}{2})\nonumber\\ 
&~~~~~~~~~~~~~~~~~~~~-(\frac{1-l(x,\gamma,\delta)}{2})\log(\frac{1-l(x,\gamma,\delta)}{2})],
\end{align}
where $l(x,\gamma,\delta)=\sqrt{1-4\lambda_0\lambda_1(1-\tanh^2 x\sin^2\gamma\cos^2\delta)}$. For $x=0.2$ and $\lambda_0=0.2$,
the minimum of weak conditional entropy is obtained for $\gamma = \pi/2$ and $\delta = 0$. 
Therefore, the super discord in the measured
state is given by $D_w(\tilde\rho_{AB}) =S(\tilde{\rho}_{B})- S(\tilde{\rho}_{AB})+S_w(A|\{Q_\eta^B(x)\}) = 0.7010$. Thus, we
have  $ D_w(\rho_{AB})- D_s(\rho_{AB}) = D_w(\tilde\rho_{AB}) = 0.7010$.

For the maximally entangled state we have $\lambda_0=\lambda_1=1/2$ and in this case we have $D_w(\rho_{AB})-D_s(\rho_{AB}) 
= -[\frac{(1-\tanh x)}{2}\log(\frac{1-\tanh x}{2}) +\frac{(1+\tanh x)}{2}\log(\frac{1+\tanh x}{2})]$. Also
$D_w(\tilde\rho_{AB})=\min_{\{\gamma,\delta\}}[-\frac{(1-\tanh x\sin\gamma\cos\delta)}{2}\log(\frac{1-\tanh x\sin\gamma\cos\delta}{2})
-\frac{(1+\tanh x\sin\gamma\cos\delta)}{2}\log(\frac{1+\tanh x\sin\gamma\cos\delta}{2})]$. The minimum occurs at
$\gamma =\pi/2$ and $\delta =0$ and at this point $D_w(\rho_{AB})-D_s(\rho_{AB})=D_w(\tilde\rho_{AB})$, showing that
indeed the super quantum discord in $\tilde\rho_{AB}$ is equal to the extra quantumness in $\rho_{AB}$.

\emph{The extra quantum correlation for the Werner state.--}
The Werner state is an admixture of a random state and a maximally entangled state, namely,
\begin{equation}
\rho_{AB} = z |\Psi^-\rangle\langle\Psi^-|+\frac{(1-z)}{4}I,
\end{equation}
where $|\Psi^-\rangle=(|01\rangle - |10\rangle)/\sqrt{2}$. For the Werner state we have $ S(A|\{\Pi^B_\psi\})
=-\frac{(1-z)}{2}\log(\frac{1-z}{2})-\frac{(1+z)}{2}\log(\frac{1+z}{2})$. In defining the conditional entropy, we have used
the computational basis. Since the Werner state is rotationally invariant therefore,  this yields the same result
for the conditional entropy for any measurement basis and hence we do not have to minimize it over all measurement bases.
The weak conditional entropy for the Werner state \cite{utm} is given by $S_w(A|\{P^B(x)\}) 
= -[\frac{(1-z\tanh x)}{2}\log(\frac{1-z\tanh x}{2}) +\frac{(1+z\tanh x)}{2}\log(\frac{1+z\tanh x}{2})].$ The Werner state
after the measurement in the computational basis is given by
\begin{align}
 \tilde{\rho}_{AB}=&\frac{(1-z)}{4}|00\rangle\langle00|+\frac{(1+z)}{4}|01\rangle\langle01|\nonumber\\
 &+\frac{(1+z)}{4}|10\rangle\langle10|+\frac{(1-z)}{4}|11\rangle\langle11|.
\end{align}
For this state, we have $S(\tilde{\rho}_{B})=1$ and $S(\tilde{\rho}_{AB})=-\frac{(1-z)}{2}\log(\frac{1-z}{4})-\frac{(1+z)}{2}
\log(\frac{1+z}{4}) = 1+S(A|\{\Pi^{B}\})$.
%{\it(We don't need this part--)
%The normal discord for this state is obtained if we calculate the conditional entropy in computational basis. As minimum of conditional entropy, which is given by
%\begin{align}
% S(\tilde{\rho}_{A|\{\Pi_\psi^B\}})=&-\frac{(1-z\cos\theta)}{2}\log(\frac{(1-z\cos\theta)}{2})\nonumber\\
% &-\frac{(1+z\cos\theta)}{2}\log(\frac{(1+z\cos\theta)}{2}),
%\end{align}
%occurs at $\theta = n\pi$, where $n=0,\pm1,..$. So for calculating weak discord in the state $\tilde{\rho}_{AB}$, one should use computational basis $(\theta = n\pi)$.
%This can be explicitly checked here.}
The weak conditional entropy for this state can be calculated using $\{Q^B_\phi(x),Q^B_\phi(-x)\}$ as a set of measurement
operators, where $Q^B_\phi(\pm x) = a(\pm x)\Pi_{\phi}+a(\mp x)\Pi_{\bar{\phi}}$ with $\{|\phi\rangle,|\bar{\phi}\rangle\}$
as an orthogonal basis. Here $|\phi\rangle=\cos\frac{\theta}{2}|0\rangle+\exp(i\Phi )\sin\frac{\theta}{2}|1\rangle$ and 
$|\bar\phi\rangle=\cos\frac{\theta}{2}|1\rangle - \exp(-i\Phi )\sin\frac{\theta}{2}|0\rangle$. Therefore, the weak conditional
entropy for the measured Werner state is given by
\begin{align}
S_w&(A|\{Q^B_\phi\})\nonumber\\ =&\min_{\theta} [-\frac{(1-z\tanh x\cos\theta)}{2}\log(\frac{1-z\tanh x\cos\theta}{2}) \nonumber\\ 
&-\frac{(1+z\tanh x\cos\theta)}{2}\log(\frac{1+z\tanh x\cos\theta}{2})].
\end{align}
The minimum value of conditional entropy occurs at $\theta = n\pi$  $(n=0,1...)$. 
For $\theta = \pi$ , we have $ S^\mathrm{min}_w(A|\{Q^{B}_\phi\}) =-[\frac{(1+z\tanh x)}{2}\log(\frac{1+z\tanh x}{2}) 
+\frac{(1+z\tanh x)}{2}\log(\frac{1+z\tanh x}{2})]=S_w(A|\{P^{B}(x)\})$. Therefore, the super discord in $\tilde{\rho}_{AB}$
is given by
\begin{align}
D_w(\tilde\rho_{AB}) &= S(\tilde{\rho}_B)-S(\tilde{\rho}_{AB})+S^\mathrm{min}_w(A|\{Q^{B}_\phi\})\nonumber\\
&= -S(A|\{\Pi^{B}\})+S_w(A|\{P^{B}(x)\})\nonumber\\
&= D_w(\rho_{AB})-D_s(\rho_{AB}) = \Delta.
\end{align}
Here the explicit value of the extra quantum correlation is given by
%\begin{align}
$\Delta = -[\frac{(1-z\tanh x)}{2}\ln(\frac{1-z\tanh x}{2})
+\frac{(1+z\tanh x)}{2}\ln(\frac{1+z\tanh x}{2})]
+[\frac{(1+z)}{2}\ln(\frac{1+z}{2})+\frac{(1-z)}{2}\ln(\frac{1-z}{2})]$.
%\end{align}
It should be noted here that the bases which minimizes conditional entropy for the original Werner state and the measured
Werner state are the same. This is consistent with our main result.

\emph{Conclusions.--}
%We have shown how weak measurements can resurrect the lost quantumness of a composite state.
One of the remarkable features of the weak measurement is that it can reveal more quantum correlation
in a bipartite state as a function of the measurement strength. This makes the quantum-classical boundary a 
dynamical one depending on the measurement strength. In this sense, the quantum correlation of a composite system looses its absolute 
meaning. Rather, it depends on how gently or strongly one perturbs a quantum system.
Surprisingly, we have shown that
%presented a general theorem showing that 
the ``extra quantum correlation'' in a state, which is quantified as the difference between the super quantum discord and the normal
discord in the state, is actually equal to the super quantum discord in the measured state after the projective measurement has been
performed on one of the subsystems in the original state.
%a state which one will get if he does a strong measurement in a basis for which normal discord is calculated.
This amounts to saying that the extra quantum correlation which appears as a consequence of the weak measurement can be extracted
even after one performs a strong projective measurement on the quantum state. This shows that the extra quantum correlation is robust against
the projective measurements and therefore we can resurrect it after a complete projective measurement, using the weak measurement.
It has been exemplified for a general pure entangled state and the Werner state.
This is a new quantum effect, which is completely counter intuitive. 
%One may tend to think that somehow the weak 
%measurement creates the quantum correlation in the post measurement state (after the 
%strong measurement). But this is not the case.
Our result shows that the extra quantum correlation in a state is a property of the system and 
the measurement strength, and it may be hinting us towards a conservation law for 
the quantum correlation for a given state. This can be stated as {\it the amount of extra quantum correlation which is destroyed by the projective measurement in the 
original state is 
equal to the amount of extra quantum correlation captured by the weak measurement in the post-measured state.}
%If there is such a conservation law then it will help in distribution of quantum correlation for a given state for  different
%applications using the same state.
%Also there are evidences that the total correlations behaves as if they were exclusively
%quantum, it will be of great importance if we can use it. 
The results of our paper provide new tools to utilize the
total correlations in a quantum state, based on the weak measurement. It may be possible to use 
the post-measured state after the local projective
measurement, which is otherwise useless, for efficient quantum information 
processing and computation tasks.
We believe that this will provide new ways of thinking about quantum correlation and making full use of it as resource in quantum information 
processing in future.

\end{document}